\documentclass[aps,pre,showkeys,showpacs,preprint]{revtex4}
\usepackage{graphicx}
\usepackage{amsmath}
\usepackage{amsfonts}
\usepackage{amssymb}
\usepackage{bm}
\usepackage{enumitem}  
\usepackage{float}
\usepackage{color, graphics}

\begin{document}
\title{On the Damped Pinney Equation from Noether Symmetry Principles}

\author{Fernando Haas\footnote{E-mail: fernando.haas@ufrgs.br. Orcid: 0000-0001-8480-6877}}
\affiliation{Instituto de F\'{\i}sica, Universidade Federal do Rio Grande do Sul, Av. Bento Gon\c{c}alves 9500, 91501-970 Porto Alegre, RS, Brasil}

\begin{abstract}
There are several versions of the damped form of the celebrated Pinney equation, which is the natural partner of the undamped time-dependent harmonic oscillator. In this work these dissipative versions of the Pinney equation are briefly reviewed. We show that Noether's theorem for the usual time-dependent harmonic oscillator as a guiding principle for derivation of the Pinney equation also works in the damped case, selecting a Noether symmetry based damped Pinney equation. The results are extended to general nonlinear damped Ermakov systems. A certain time-rescaling always allows to remove the damping from the final equations. 
\end{abstract}



\maketitle

\section{Introduction}

The Pinney equation \cite{Pinney} is the natural partner of the time-dependent harmonic oscillator (TDHO) equation, as was recognized from its relevance for the construction of the Ermakov invariant \cite{Ermakov}. It is reasonable to ask what could some distinguished dissipative version of such model, if existing. Indeed, the subject called attention in the literature and there are some alternative versions of the {\it damped Pinney equation}. The present work strictly follows the Noether theorem as a guiding principle for the derivation of the Pinney equation, both in damped and undamped cases. The results found suggest an extension to general dissipative nonlinear Ermakov systems, as we shall show.   

The proposal of symmetry principles for time-dependent dissipative systems certainly is not new. For instance, Cari\~nena and de Lucas considered Lie systems for certain dissipative Ermakov systems \cite{Carinena2009}.  Brazier and Leach \cite{Leach} have applied Noether's theorem for the Lagrangians of the kind
\begin{equation}
\label{e1}
L(q,\dot q, t) = \frac{m(t)}{2}(\dot q^2 - \omega^{2}(t)q^2) \,,
\end{equation}
and nonlinear generalizations, see also \cite{Gonzalez}, and derived a damped version of the Pinney equation, to be reviewed here. The quadratic potential in Eq. (\ref{e1}) has been also treated elsewhere, as in \cite{pro1}--\cite{simic}. 

Some different models consider, for instance, approximate solutions for the simplest damped Pinney equation, obtained just adding a damping term linear in the velocity \cite{Haas}. Dissipative harmonic oscillator systems with a quadratic dependence on the velocity together with non-standard Lagrangians can be also enumerated \cite{Musil1, Musil2, Qian}. Generalized damped Ermakov systems have been derived by means of Sundman transformations \cite{Guha}. Dissipative Ermakov-Milne-Pinney systems with Li\'enard type equations can be also derived, by means of nonlocal transformations \cite{Gu}. Integrable equations with Ermakov-Pinney nonlinearities and Chiellini damping have been found \cite{Mancas}, also in the treatment of barotropic FRW cosmological models \cite{manc}.


The present work considers only Noether point symmetries as a guiding principle for derivation of the dissipative Pinney equation and the corresponding invariants. Obviously this does not imply any criticism to the available alternative versions of damped Pinney and Ermakov equations. The presented treatment follows a progressive route, starting from the simplest to the more complex cases. Section 2 presents the undamped case and basic methods. Section 3 details the physically important case of a constant linear in velocity damping term. Section 4 unveil the more general case of a damping term linear in the velocity and also admitting time dependence. The known results in the literature are then reviewed. Section 5 extends the previous results to general nonlinear damped Ermakov systems. It is also shown that the damping term in the symmetry-generated damped Pinney and Ermakov equations and dissipative TDHO equation can be always removed by means of a rescaling of time. Section 6 is reserved for the conclusions.

\section{Pinney equation from Noether symmetry of the time-dependent harmonic oscillator}

The TDHO equation 
\begin{equation}
\label{tdho}
    \ddot q + \omega^{2}(t)q = 0 
\end{equation}
is derivable from the Lagrangian 
\begin{equation}
\label{ll}
    L = \frac{1}{2}(\dot q^2 - \omega^{2}(t)q^2) \,.
\end{equation}
The existence of a variational principle allows the search for first integrals following Noether's theorem.

According to Noether's theorem, the invariance of the action functional (up to addition of a numerical constant) can be associated to a constant of motion. 
For a Lagrangian $L = L(q,\dot q,t)$,  consider a general point symmetry generator 
\begin{equation}
G = \tau\frac{\partial}{\partial t} + \eta\frac{\partial}{\partial q} \,,   
\end{equation}
where $\tau = \tau(q,t), \, \eta = \eta(q,t)$. The corresponding  generator of the first extended group is then 
\begin{equation}
G^{(1)} = G + (\dot\eta - \dot\tau\dot q)\frac{\partial}{\partial\dot q} \,,
\end{equation}
where 
\begin{equation}
\dot\tau = \frac{\partial\tau}{\partial t} + \dot q\frac{\partial \tau}{\partial q} \,,  \quad \dot{\eta} = \frac{\partial\eta}{\partial t} + \dot q\frac{\partial\eta}{\partial q} \,.
\end{equation}
The Noether symmetry condition is
\begin{equation}
\label{nsc}
G^{(1)}[L] + \dot\tau L = \dot F = \frac{\partial F}{\partial t} + \dot q\frac{\partial F}{\partial q} \,,
\end{equation}
where for a point transformation one has a gauge function $ F =  F(q,t)$. The associated Noether invariant is 
\begin{equation}
\label{ni}
I = \tau\left(\dot q\cdot\frac{\partial L}{\partial\dot q} - L\right)  -   \eta\frac{\partial L}{\partial\dot q} +  F \,, \quad \frac{dI}{dt} = 0 \,. 
\end{equation}

Inserting the Lagrangian from Eq. (\ref{ll}) into the symmetry condition (\ref{nsc}) yields a third-order polynomial in the velocity, whose monomials corresponding to different powers of $\dot q$ should all be zero. Solving the resulting equations is a simple subject, yielding 
\begin{equation}
    \tau = \tau(t) \,, \quad \eta = \frac{\dot\tau q}{2} - \gamma(t) \,, \quad F = \frac{\ddot\tau q^2}{4} - \dot\gamma q \,,
\end{equation}
where 
\begin{eqnarray}
\label{n1}
    \dddot\tau &+& 4\omega^2\dot\tau + 4\omega\dot\omega\tau = 0 \,,\\
    \label{n2}
    \ddot\gamma &+& \omega^2\gamma = 0 \,.
\end{eqnarray}
Details on the procedure and the explicit realization of the symmetry algebra can be found in \cite{gra}, with application to graphene models. From Eq. (\ref{ni}) the resulting Noether invariant is
\begin{equation}
\label{n3}
    I = \frac{1}{2}\left[\tau\dot q^2 - \dot\tau q\dot q + (\ddot\tau + 2\omega^2\tau)\frac{q^2}{2}\right] + \gamma\dot q - \dot\gamma q \,.
\end{equation}

At this moment one can set $\tau = \rho^2$, where $\rho = \rho(t)$ is a function to be determined. Integrating once the linear homogeneous third-order equation (\ref{n1}) yields the Pinney equation
\begin{equation}
\label{pi}
    \ddot\rho + \omega^2\rho = \frac{k}{\rho^3} \,,
\end{equation}
where $k$ is the integration constant, a first integral for Eq. (\ref{n1}) in this context \cite{Andri}. A simple algebra allows to rewrite the Noether invariant (\ref{n3}) as 
\begin{equation}
\label{erm}
    I = \frac{1}{2}(\rho\dot q - \dot\rho q)^2 + \frac{k}{2}\left(\frac{q}{\rho}\right)^2 \,,
\end{equation}
after setting $\gamma = 0$ disregarding the Wronskian part. The first integral in Eq. (\ref{n3}) is known as Ermakov invariant \cite{Ermakov} and appears in many areas. In this sense, the partner equation (\ref{pi}) for the TDHO equation follows from Noether symmetry. This provides a symmetry principle for the derivation of  Pinney's equation. It can be remarked that the invariant (\ref{erm}) can be directly deduced after eliminating $\omega^2$ between Eqs. (\ref{tdho}) and (\ref{erm}), without reference to symmetry invariance \cite{Pinney}, see also e.g. \cite{RR}.  Moreover, the Ermakov invariant is invariant under the transformation group, namely $G^{(1)}[I] = 0$ as readily found - this is a general property, the form invariance of the Noether invariant under Noether symmetry \cite{Sarlet}. Also notice that actually Eq. (\ref{n3}) contains 5 invariants, corresponding to the linearly independent solutions of the system (\ref{n1})-(\ref{n2}). Moreover, the group of Noether point symmetries  is a subgroup of the full point symmetry group of the TDHO \cite{LL}. A recent review on Ermakov systems and their symmetries appears in \cite{cerv}.


\section{Pinney equation from Noether symmetry of the damped time-dependent harmonic oscillator}

The damped TDHO equation 
\begin{equation}
\label{xtdho}
    \ddot q + \lambda\dot q + \omega^{2}(t)q = 0 
\end{equation}
where $\lambda > 0$ is a numerical constant, 
is derivable from the Caldirola-Kanai \cite{c, k} Lagrangian 
\begin{equation}
\label{Xll}
    L = \frac{e^{\lambda t}}{2}(\dot q^2 - \omega^{2}(t)q^2) \,.
\end{equation}

Inserting the Lagrangian from Eq. (\ref{Xll}) into the symmetry condition (\ref{nsc}) yields a third-order polynomial in the velocity $\dot q$, whose monomials corresponding to different powers of $\dot q$ should all be zero. Solving the resulting equations yields 
\begin{equation}
    \tau = \tau(t) \,, \quad \eta = (\dot\tau - \lambda\tau)\frac{q}{2} - \gamma(t) \,, \quad F = e^{\lambda t}(\ddot\tau - \lambda\dot\tau)\frac{q^2}{4} - e^{\lambda t}\dot\gamma q \,,
\end{equation}
where 
\begin{eqnarray}
\label{xn1}
    \dddot\tau &+& (4\omega^2 - \lambda^2)\dot\tau + 4\omega\dot\omega\tau = 0 \,,\\
    \label{xn2}
    \ddot\gamma &+& \lambda\dot\gamma + \omega^2\gamma = 0 \,.
\end{eqnarray}
From Eq. (\ref{ni}) the Noether invariant is
\begin{equation}
\label{xn3}
    I = \frac{e^{\lambda t}}{2}\left[\tau\dot q^2 - (\dot\tau - \lambda\tau) q\dot q + (\ddot\tau - \lambda\dot\tau + 2\omega^2\tau)\frac{q^2}{2}\right] + e^{\lambda t}(\gamma\dot q - \dot\gamma q) \,.
\end{equation}
Obviously setting $\lambda = 0$ recovers the undamped case results.

At this moment we can set $\tau = \rho^2 e^{\lambda t}$, where $\rho = \rho(t)$ is a function to be determined. Integrating once the linear homogeneous third-order equation (\ref{xn1}) yields 
\begin{equation}
\label{xpi}
    \ddot\rho + \lambda\dot\rho + \omega^2\rho = \frac{k\,e^{-2\lambda t}}{\rho^3} \,,
\end{equation}
where $k$ is an integration constant. Eq. (\ref{xpi}) is the legitimate damped Pinney equation as the natural partner of the damped TDHO equation, in the context of Noether symmetry principles.

A simple algebra allows to rewrite the Noether invariant (\ref{xn3}) as 
\begin{equation}
\label{xerm}
    I = \frac{e^{2\lambda t}}{2}(\rho\dot q - \dot\rho q)^2 + \frac{k}{2}\left(\frac{q}{\rho}\right)^2 \,,
\end{equation}
after setting $\gamma = 0$ disregarding the Wronskian part.


Equations (\ref{xpi}) and (\ref{xerm}) provide damped versions of the Pinney equation and Ermakov invariant, in the simplest case of a constant (in time) linear in velocity damping. Moreover, the Noether and Ermakov invariants in Eqs. (\ref{xn3}) and (\ref{xerm}) are both invariant under the transformation group, namely $G^{(1)}[I] = 0$ as readily found. Notice (\ref{xn3}) actually contains 5 invariants, corresponding to 5 linearly independent solutions of the system (\ref{xn1})-(\ref{xn2}). It can be verified, that $G^{(1)}[I] = 0$ in the same way as in the undamped case.

In the same sense as for the undamped case, the Noether invariant can be also deduced from eliminating $\omega^2$ between Eqs. (\ref{xtdho}) and (\ref{xpi}), without reference to symmetry invariance. After a short manipulation this procedure yields 
\begin{equation}
    e^{2\lambda t} \left[(\rho\dot q - \dot\rho q)(\rho\ddot q - \ddot\rho q) + \lambda(\rho\dot q - \dot\rho q)^2\right]  + \frac{k\,q (\rho\dot q - \dot\rho q)}{\rho^3} = 0 \,,
    \end{equation}
which can be integrated producing the invariant in Eq. (\ref{xerm}), set $\lambda = 0$ for the undamped limit. Notice the time-dependence of the nonlinear term is essential in the damped Pinney equation (\ref{xpi}), for the existence of the constant of motion.

\section{Time-dependent linear in velocity damping}

Starting from the Lagrangian (\ref{e1}) yields 
\begin{equation}
\label{lea}
    \ddot q + \lambda(t)\dot q + \omega^{2}(t)q = 0 \,, 
\end{equation}
where $\lambda = \lambda(t) = \dot{m}/m$ is a time-dependent damping coefficient, generalizing the previous models. This case can be interpreted as a TDHO with a time-dependent mass - the previous section corresponds to a exponentially growing mass. 

The analysis of Noether symmetries for this problem is well known \cite{Leach}. In terms of the present notation the results are
\begin{equation}
    \tau = \tau(t) \,, \quad \eta = (\dot\tau - \lambda\tau)\frac{q}{2} - \gamma(t) \,, \quad F = m(t)\left[(\ddot\tau - \lambda\dot\tau - \dot\lambda\tau)\frac{q^2}{4} - \dot\gamma q\right] \,,
\end{equation}
where 
\begin{eqnarray}
\label{xxn1}
    \dddot\tau &+& (4\omega^2 - \lambda^2 - 2\dot\lambda)\dot\tau + (4\omega\dot\omega - \ddot\lambda - \lambda\dot\lambda)\tau = 0 \,,\\
    \label{xxn2}
    \ddot\gamma &+& \lambda\dot\gamma + \omega^2\gamma = 0 \,.
\end{eqnarray}
The Noether invariant is
\begin{equation}
\label{xxn3}
    I = \frac{m(t)}{2}\left[\tau\dot q^2 - (\dot\tau - \lambda\tau) q\dot q + (\ddot\tau - \lambda\dot\tau - \dot\lambda\tau + 2\omega^2\tau)\frac{q^2}{2}\right] + m(t)(\gamma\dot q - \dot\gamma q) \,.
\end{equation}

Now setting $\tau = m \rho^2$ where $\rho = \rho(t)$ is a function to be determined gives
\begin{equation}
\label{xxpi}
    \ddot\rho + \lambda\dot\rho + \omega^2\rho = \frac{k}{m^2\rho^3} \,, \quad \lambda = \frac{\dot m}{m} \,,
\end{equation}
where $k$ is a numerical constant found reducing the order of Eq. (\ref{xxn1}). Setting $\gamma = 0$ disregarding the Wronskian part, the first integral in Eq. (\ref{xxn3}) assumes the compact form  
\begin{equation}
\label{xxerm}
    I = \frac{m^2}{2}(\rho\dot q - \dot\rho q)^2 + \frac{k}{2}\left(\frac{q}{\rho}\right)^2 \,,
\end{equation}

Similar remarks from the previous sections apply here, including the Ermakov invariant (\ref{xxerm}) being derivable eliminating the frequency between Eqs. (\ref{lea}) and (\ref{xxpi}). The previous results are all re-obtained setting either $m = 1$ (undamped TDHO) or $m = \exp(\lambda t)$ (for a constant damping coefficient $\lambda$).

Equation (\ref{xxpi}) is the proper Pinney equation in the case of a time-dependent linear in velocity damped TDHO, derived from Noether symmetry. It was also derived in this context by Brazier and Leach \cite{Leach}. This damped Pinney equation and the invariant (\ref{xxerm}) were found using different alternative approaches in many opportunities. For instance in the treatment of Berry phases \cite{Maa}, parametric anharmonic oscillators \cite{Galle}, energy exchanges in the 
dissipative TDHO \cite{guasti}, from a canonical transformation method for the variable-mass TDHO \cite{pedro}  and quantum harmonic oscillators with time-dependent mass and frequency from a perturbative potential method \cite{dantas}. It has been also identified from perturbations of Lagrangian systems based on the preservation of subalgebras of Noether symmetries \cite{campo}.

\section{Nonlinear damped Ermakov systems - discussion and generalization}

The examination of the previous damped TDHO and damped Pinney equations suggests a nonlinear extension, based on the general Ermakov (or Ray-Reid) system \cite{RR}, given by 
\begin{eqnarray}
\label{rr1}
    \ddot q + \lambda(t)\dot q + \omega^{2}(t)q &=& \frac{1}{m^{2}}\,\frac{1}{\rho q^2}\,f\left(\frac{\rho}{q}\right) \,, \\
\label{rr2}
    \ddot\rho + \lambda(t)\dot\rho + \omega^2(t)\rho &=& \frac{1}{m^{2}}\,\frac{1}{q\rho^2}\,g\left(\frac{q}{\rho}\right) \,,
\end{eqnarray}
where $m = m(t)$, $\lambda = \lambda(t) = \dot m/m$ and $f, g$ are arbitrary functions of the indicated arguments. 

Eliminating the frequency from Eqs. (\ref{rr1})-(\ref{rr2}) and proceeding as before provides the invariant
\begin{equation}
\label{rr3}
    I = \frac{m^2}{2}(\rho\dot q - \dot\rho q)^2 + \int^{\rho/q}f(s)ds + \int^{q/\rho}g(s)ds \,, 
\end{equation}
which is a constant of motion, $dI/dt = 0$, as can be verified. Setting $m = 1$ reduces to the traditional general Ermakov system \cite{RR}. Otherwise we have a realization of a damped general nonlinear Ermakov system. In spite of its simplicity, the system (\ref{rr1})-(\ref{rr2}) in connection with the invariant (\ref{rr3}) is apparently new, to the best of our knowledge. Unlike for the derivation of damped Pinney equations, the present derivation is not based on symmetry principles but rather on the direct manipulation of the dynamical equations. Judicious choices of $m, f$ and $g$ encompasses all the previous models.

It can be observed that a time-rescaling allows to eliminate the linear in velocity damping term. Defining a new time variable $T = T(t)$ such that 
\begin{equation}
\label{res}
    \dot{T} = 1/m
\end{equation}
converts Eqs. (\ref{x1})-(\ref{x2}) into
\begin{eqnarray}
\label{x1}
    q'' + \Omega^{2}(t)q &=& \frac{1}{\rho q^2}\,f\left(\frac{\rho}{q}\right) \,, \\
\label{x2}
    \rho'' + \Omega^2(t)\rho &=& \frac{1}{q\rho^2}\,g\left(\frac{q}{\rho}\right) \,,
\end{eqnarray}
where $\Omega^2 = m^2 \omega^2$ and a prime denotes derivative with respect to the new time variable $T$. The Ermakov invariant becomes
\begin{equation}
\label{x3}
    I = \frac{1}{2}(\rho q' - \rho' q)^2 + \int^{\rho/q}f(s)ds + \int^{q/\rho}g(s)ds \,, 
\end{equation}
all as if there was no damping at all, coming back to the traditional formulation \cite{RR}. The time-rescaling applies to all Noether symmetry based Pinney equations and damped TDHOs considered here. In the case of a constant damping coefficient, this damping removal was done in the treatment of dissipative Bose-Einstein condensates \cite{schuch}.

In practice, however, the formal removal of damping can be not so much useful. For instance, consider a constant $\omega = \omega_0$ and a constant damping coefficient so that $\dot\lambda = 0$. In this case, one has $m = \exp(\lambda t)$, $\dot T = \exp(-\lambda t)$ and finally
\begin{equation}
\label{T}
T = \frac{1}{\lambda}\,(1 - e^{-\lambda t}) \,,
\end{equation}
setting $T(0) = 0$ without loss of generality. Therefore the rescaled angular frequency comes from 
\begin{equation}
    \Omega^2 = e^{2\lambda t}\omega_{0}^2 = \frac{\omega_{0}^2}{(1- \lambda T)^2} \,,
\end{equation}
which is asymptotically singular as $T \rightarrow 1/\lambda$. Therefore the removal of damping comes at the expense of the introduction of singular terms (in time) in the equations of motion. The behavior of the new time from Eq. (\ref{T}) is shown in Fig. 1. 

\begin{figure}[H]
\includegraphics[width=10.5 cm]{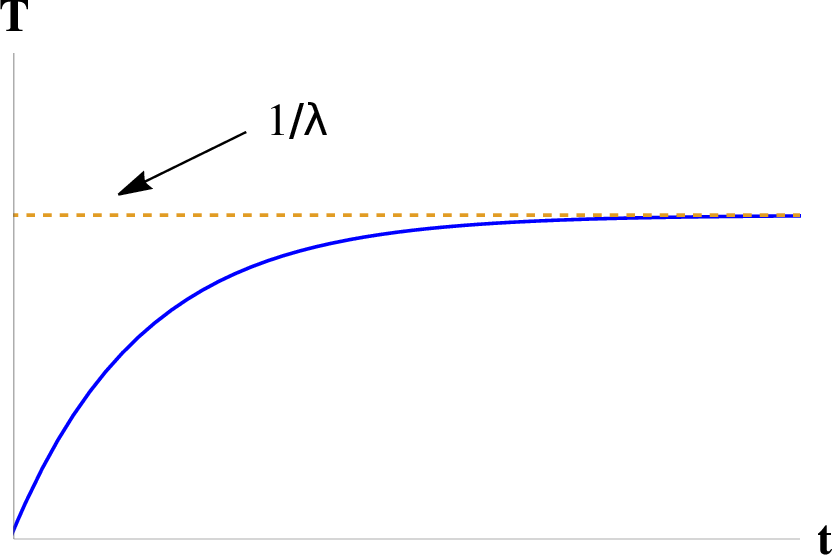}
\caption{Generic behavior of the new time $T$ from Eq. (\ref{T}). Dashed: the asymptotic limit $T \rightarrow 1/\lambda$. \label{fig1}}
\end{figure}   
\unskip

The presence of the factor $1/m^2$ on the right-hand side of Eqs. (\ref{rr1}--\ref{rr2}) is essential for the existence of the Ermakov invariant. Direct addition of linear in velocity damping terms would gives the system
\begin{eqnarray}
\label{rrr1}
    \ddot q + \lambda(t)\dot q + \omega^{2}(t)q &=& \frac{1}{\rho q^2}\,f\left(\frac{\rho}{q}\right) \,, \\
\label{rrr2}
    \ddot\rho + \lambda(t)\dot\rho + \omega^2(t)\rho &=& \frac{1}{q\rho^2}\,g\left(\frac{q}{\rho}\right) \,,
\end{eqnarray}
After eliminating the frequency between Eqs. (\ref{rrr1})--(\ref{rrr2}) and defining 
\begin{equation}
\label{rrr3}
    \tilde{I} = \frac{m^2}{2}\left[(\rho\dot q - \dot\rho q)^2 + \int^{\rho/q}f(s)ds + \int^{q/\rho}g(s)ds\right] \,, \quad \lambda = \frac{\dot m}{m}
\end{equation}
just yields
\begin{equation}
\frac{d\tilde{I}}{dt} = - 2 m\dot{m} \, \left(\int^{\rho/q}f(s)ds + \int^{q/\rho}g(s)ds\right) \,,
\end{equation}
which is not a conservation law in general. Following Nassar \cite{Nassar}, Eqs. (\ref{rrr1})-(\ref{rrr2}) can be termed a damped non-Ermakov system, while Eqs. (\ref{rr1})-(\ref{rr2}) can be termed a damped Ermakov system. 


As a final, strong generalization, notice that the functions $\omega, m, \lambda$ in Eqs. (\ref{rr1})-(\ref{rr2}) do not need to be restricted to be functions of time only, without prejudice of the existence of the first integral in Eq. (\ref{rr3}), as long as $\lambda = \dot{m}/m$. Indeed, the derivation of the invariant relies on the elimination of $\omega^2$ between the dynamical equations and the identification of the appropriate integrating factor. Consider the dynamical system
\begin{eqnarray}
\label{m1}
    \ddot q + \lambda\dot q + \omega^{2}q &=& \frac{1}{m^{2}}\,\frac{1}{\rho q^2}\,f\left(\frac{\rho}{q}\right) \,, \\
\label{m2}
    \ddot\rho + \lambda\dot\rho + \omega^2\rho &=& \frac{1}{m^{2}}\,\frac{1}{q\rho^2}\,g\left(\frac{q}{\rho}\right) \,,
\end{eqnarray}
without specific details about the functions $\lambda, \omega$ and $m$. The elimination procedure then yields 
\begin{equation}
    \frac{dI}{dt} + m^2 \left(\lambda - \frac{\dot m}{m}\right) (\rho\dot q - \dot\rho q)^2 = 0 \,,
\end{equation}
%
where $I$ is the generalized Ermakov invariant in Eq. (\ref{rr3}), which is then a constant of motion provided $\lambda = \dot{m}/m$. 

As an illustration, suppose 
\begin{equation}
\label{mw}
    m = m(q,\rho,t) \,, \quad \omega = \omega(q,\rho,t) \,,
\end{equation}
so that 
\begin{equation}
\label{lam}
    \lambda = \frac{1}{m}\left(\frac{\partial m}{\partial q}\dot q + \frac{\partial m}{\partial \rho}\dot\rho + \frac{\partial m}{\partial t}\right) \,.
\end{equation}
 As demonstrated, the system (\ref{m1})-(\ref{m2}) admits the invariant (\ref{rr3}), irrespective of much details about $m, \omega$ besides Eq. (\ref{mw}), provided $\lambda$ respects Eq. (\ref{lam}). In this particular case the damping terms can have a quadratic dependence on the velocities $\dot{q}, \dot\rho$.
 
 To sum up, in principle one has a new, large class of damped generalized Ermakov systems, which remains to be explored in more detail.

\section{Conclusions}

The damped Pinney equation was discussed in detail, from the point of view of Noether symmetry derivations. The general Noether symmetry based damped Pinney equation (\ref{xxpi}) was definitely identified, together with the associated Ermakov invariant (\ref{xxerm}), with a review on the existing literature about it. Following the route of a progressive analysis - from the undamped TDHO, to the damped TDHO with a constant damping coefficient, to the damped TDHO with time-dependent linear in velocity term - it is natural to consider the extension of the results to damped general nonlinear Ermakov systems. A new, large class of damped nonlinear Ermakov systems was then proposed. The TDHO and Pinney equations are ubiquitous in physics, engineering and beyond, for instance in plasma physics, gravitation and quantum optics (see e.g. Ref. \cite{dantas} and references therein). Hence for the more realistic scenario where some damping mechanism is present, the damped Pinney equation is expected to be useful, as in the case of generalized dissipative cosmological models \cite{manc}. Moreover, the new class of damped nonlinear Ermakov system can have applications for example in the case of a pendulum whose mass depends not only on time, but also on space. It would be important to unveil the properties of this new system in more detail, for instance regarding its integrability or chaos features, in spite of the existence of the Ermakov invariant (\ref{rr3}), which is not a sufficient condition for complete integrability. Finally, the removal of the damping in some specific cases by a re-definition of the time variable was critically discussed.

\acknowledgments
The author acknowledges the support by Con\-se\-lho Na\-cio\-nal de De\-sen\-vol\-vi\-men\-to Cien\-t\'{\i}\-fi\-co e Tec\-no\-l\'o\-gi\-co
(CNPq).  


\end{document}